# Power Flow Analysis Using Graph based Combination of Iterative Methods and Vertex Contraction Approach

Chen Yuan, *Member, IEEE*, Guangyi Liu, *Senior Member, IEEE*, Renchang Dai, *Senior Member, IEEE*, Zhiwei Wang

*Abstract*—Compared with relational database (RDB), graph database (GDB) is a more intuitive expression of the real world. Each node in the GDB is a both storage and logic unit. Since it is connected to its neighboring nodes through edges, and its neighboring information could be easily obtained in one-step graph traversal. It is able to conduct local computation independently and all nodes can do their local work in parallel. Then the whole system can be maximally analyzed and assessed in parallel to largely improve the computation performance without sacrificing the precision of final results. This paper firstly introduces graph database, power system graph modeling and potential graph computing applications in power systems. Two iterative methods based on graph database and PageRank are presented and their convergence are discussed. Vertex contraction is proposed to improve the performance by eliminating zero-impedance branch. A combination of the two iterative methods is proposed to make use of their advantages. Testing results based on a provincial 1425-bus system demonstrate that the proposed comprehensive approach is a good candidate for power flow analysis.

*Index Terms*— Graph database, high-performance computing, iterative method, power flow, vertex contraction.

## I. Introduction

POWER flow plays the fundamental and critical role in the power system analysis. It acts as the basic function for most of applications in power system Energy Management Systems (EMS), like state estimation, security constrained unit commitment, system security assessment, "N-1" contingency analysis and transient stability. In modern power grids, high penetrations of renewable energy resources, distributed generators, power electronic devices, energy storage system, community microgrids and HVDC transmission are inevitable

This work was supported by State Grid Corporation technology project SGRIJSKJ(2016)800.
C. Yuan, G. Liu, R. Dai, and Z. Wang are with the Global Energy Interconnection Research Institute North America (chen.yuan@geirina.net, guangyi.liu@geirina.net, renchang.dai@geirina.net).

and making power systems more complicated and unpredictable with frequent fluctuations and intermittence. So, the operating conditions should consider the addition of renewable energy, distributed generation, and energy storage at the transmission and distribution levels, as well as load demand changing characteristics.

The transition from conventional power grids to modern power grids has been accelerated by public and private investments. As presented in Fig. 1, California ISO depicted California netload curves ranging from 2012 to 2020 and each curve represents a day in the month of March [1]. The "duck curve" illustrates the emerging conditions, including short, steep ramps and over-generation risk. At each timepoint, netload equals to the value of load demand minus renewable generation. From this figure, it clearly shows that, from 2012 to 2020, since the increasing integrations of solar energy and the plentiful solar irradiation in the daytime in the state of California, the netload curve largely dips in the midday. However, during the period of sunset, which is also the after-hours in the early evening, solar generation decreases and the energy demand spikes. It results in a rapid ramp in the "duck curve", as shown in Fig. 1, approximate 13,000 MW netload increment in three hours, and requires increased system flexibility to meet challenges with such steep netload ramps, over-generation risks, reliability issues [2], and even unexpected contingencies [3], [4]. If using the commercial EMS, although data could be sent from supervisory control and data acquisition (SCADA) every 5 seconds, computation analysis describing the system status and assessing possible contingencies can take minutes to hours. Therefore, operators can only monitor the system state and make decisions in minutes. In a severe event, the power grid could transition to an unstable state within seconds, making it extremely challenging for operators to respond without feasible decision support and planning analysis based on fast state estimation and power flow analysis.

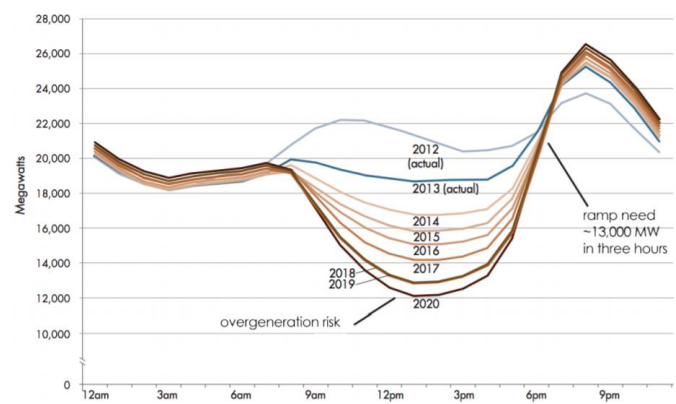

Fig. 1 California ISO projected electricity supply

A fast, even faster than real-time, computing algorithm for quickly and efficiently estimating system states and solving power flow has a profound influence on EMS performance. Parallel computing is one of promising methods to improve computation efficiency. However, the state of art of power flow analysis does not effectively make use of the parallel capability, since the relational database and computation algorithm used for existing power flow analysis were not specifically designed for parallel computing. With the fast development of computing technology and graph theory based applications, graph based high performance computation, graph computing, is a feasible option for high-performance parallel computing [5], since it was developed to deal with distributed storage and parallel computing in big data analysis, and applicable to solve complicated scenarios with iterations [6].

Commercial EMS mostly use the fast-decoupled method and Newton-Raphson method to do power flow analysis, because of its good convergence-rate. The algorithm uses a two-dimensional sparse matrix, admittance matrix, to represent the system topology. Algorithms for transmission system power flow analysis, including technologies of sparse matrix [7], sparse vector [8] and node ordering [9], have been well studied. With the evolution of software systems and hardware configurations in parallel computing, the external conditions of the power flow analysis in large-scale systems become mature. Reference [10] used distributed computation technology to implement parallel computation of power flow. Besides, GPU based parallel computing was introduced and applied to power flow calculation [11], [12]. On the other hand, our previous research works have investigated the feasibility and the high performance of graph database in power system energy management systems (EMS) applications, like CIM/E based network topology processing, power grid modeling, state estimation, "N-1" contingency analysis, and security constrained economic dispatch [13]–[17].

In this paper, a graph-based combination of iterative methods and vertex contraction approach is proposed for power flow analysis. Graph is an intuitive way to represent the world. Each node in the graph database (GDB) is a storage and logic unit, which is capable of independently and locally conducting its computation. This is because each node could easily acquire its neighboring information via one-step graph traversal. This paper will first provide a brief introduction of GDB, power system graph modeling and graph computing applications in power systems. The two selected iterative methods and vertex contraction approach for processing zero-impedance branch are presented using graph computing and then they are merged into a graph based hybrid method to make use of each algorithm's advantages. The speed and accuracy performance of the proposed method is tested and well-demonstrated with a provincial system, FJ-1425 system.

This paper is organized as follows: graph database and graph computing will be briefly introduced in Section II, including its applications in power systems. Then the proposed algorithm for power flow analysis using graph database is well elaborated in Section III. Section IV verifies the proposed algorithm accuracy using two practical systems and demonstrates its high computation performance. The potential application scenario for the proposed method will be discussed in Section V. At last, the paper is concluded in Section V and future work is also presented in this section.

## II. GRAPH DATABASE AND GRAPH COMPUTING

### A. Graph Database

Graph is a data structure modeling pairwise relations between objects in a network. In mathematics, a graph is represented as $G=(V, E)$, in which $V$ is a set of vertices, representing objects in the depicted system, and the set of edges in the graph is denoted as $E$, expressing how these vertices relate to each other. Each edge is denoted by $e=(i, j)$ in $E$, where we refer to $i$ and $j$ in $V$ as head and tail of the edge $e$, respectively.

In a GDB, each node is independent to others and capable of conducting the local computation. GDB uses graph structures for semantic queries to represent and store data in vertices and edges. Data in the GDB store are directly linked together and easily retrieved in one graph operation. So, compared with relational database (RDB), which is based on the relational model to store data, GDB permits data management in its intuitive structure. Fig. 2 presents a data storage comparison between relational database and graph database using the IEEE 5-bus system. RDB needs redundant storage to store common attributes in bridge tables to provide JOIN functions. Besides, the data query is complicated through JOIN operations and the time cost exponentially increases with the increase of the data set size. However, GDB is a direct expression of a real system. JOIN operation is no more needed, and data information are stored as attributes in corresponding nodes and edges. For example, in Fig. 2, the 5-bus system keeps the same topology in its GDB and system information are respectively assigned to vertices and edges. Then, the operations related to data query are becoming more convenient via graph operations. On the other hand, a testing on an open-source GDB management system, Neo4j, against a widely used RDB management system, MySQL, shows that the overall performance of data search in Neo4j is much better than MySQL [18].

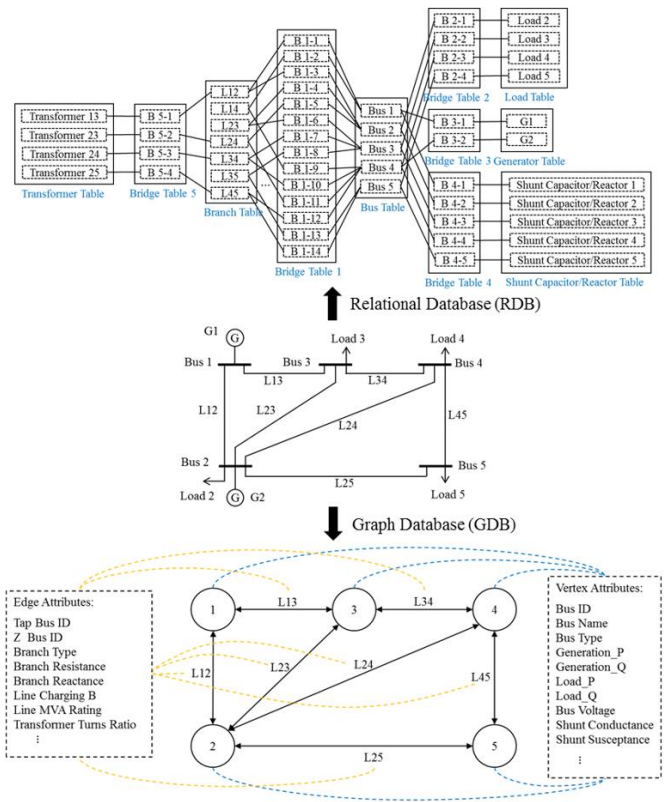

Fig. 2. Comparison between RDB and GDB in power systems modeling

### B. Power System Graph Modeling

In this section, this paper explores and demonstrates the GDB modeling in power systems. An example of mapping between graph and power system is presented in Fig. 3. A 6-bus power system is converted to a graph with the same structure, containing 6 vertices and 7 undirected edges. For a n-bus power system, its admittance matrix is a $n \times n$ symmetrical matrix. It not only stores the nodal admittance and line admittance of each bus, but also represents the topology structure of the power system. That is also a reason that the admittance matrix is very sparse. The admittance matrix is even more sparse when the system's scale is larger. Furthermore, each diagonal element approximately equals the negative sum of off-diagonal elements in the corresponding row/column, and the small mismatch, if any, is caused by the shunt admittance and the transformer turns ratio. In other words, the sum of each row/column is close to zero, or exactly zero if no shunt admittance and no transformer exists. In the mathematical field of graph theory, the Laplacian matrix, also called admittance matrix, is a matrix representation of a graph. It is equal to the graph's degree matrix minus the adjacency matrix. For an undirected graph, which is applicable to power systems, the Laplacian matrix is symmetrical, and the summation of elements in each row/column is zero. So, the power system and the undirected graph are closely mapping to each other, indicating the feasible applications of GDB into power systems.

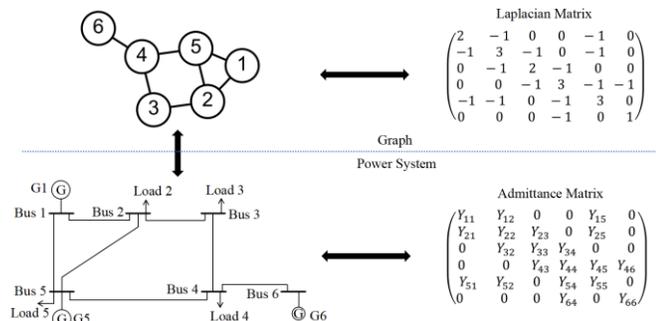

Fig. 3. Mapping between graph and power system

### C. Node-based Parallel Graph Computing and Its Applications in Power System Analysis

*1) Node-Based Parallel Computing*: In graph computing, each node is independent to others and capable of conducting the local computation. In the mode of all-node-synchronization, nodes are all activated at the same time. Then the node-based graph operation is implemented in parallel to save computation time and improve the computation efficiency. This paper proposes to employ graph computing for power flow analysis. Take the admittance matrix as an example, off-diagonal elements are locally and independently calculated based on the admittance attributes of the corresponding edges, and each diagonal element is calculated by summing the admittance attributes of the node and its connected edges. So, the admittance matrix can be formulated in parallel using one graph operation. Other examples of node-based parallel computation in power flow analysis are power injection update, system states mismatch, convergence check, branch power flow calculation, etc.

*2) MapReduce and Bulk Synchronous Parallel*: MapReduce and bulk synchronous parallel (BSP) are two major parallel computation models. BSP is a bridging model to design parallel algorithms. It contains components who are capable of local memory transactions, a network that communicates messages between components, and a facility allows for synchronization. As shown in Fig. 4, within the graph processing engine (GPE), the master processor assigns tasks to worker processors per the CPU resources, data partitions and job request. Each worker focuses on its local computation, communicates with other workers, and outputs results to barrier synchronization. The whole process is implemented in BSP. For each worker, it employs MapReduce scheme to do local logic and algebraic operation in parallel. MapReduce is a framework of processing massive datasets in form of <key; value> pairs and plays a prominent role in parallel computing. It includes two phases, map phase, performing local data processing in parallel, and reduce phase, processing output data per key in parallel. Below is the MapReduce programming mechanism in graph computing. Using SELECT syntax, MapReduce processes are generated for selected nodes. Each node's MapReduce is processed in parallel. Beginning from ACCUM syntax, map phase starts to do edges operations for the corresponding node. In the POST-ACCUM, reduce phase updates and aggregates results for each node.

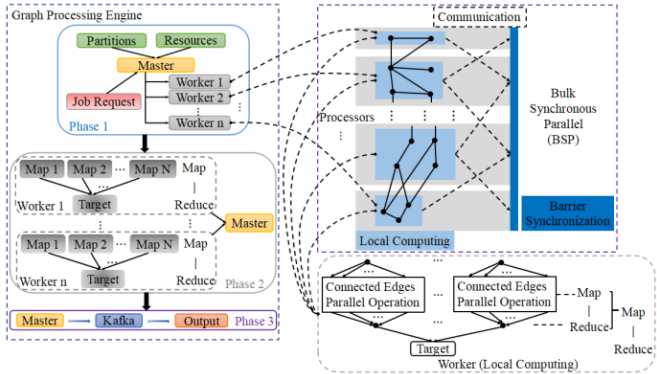

Fig. 4. MapReduce and BSP inside graph processing engine

| MapReduce in Graph Computing |
|---|
| 1. Initialize T0 = {all nodes}; |
| 2. T1 = SELECT s FROM T0: s-(edges:e)−>t |
|    // Start MapReduce processes for selected nodes |
| 3.        ACCUM |
| 4.        [edge operations]    // map phase for each selected node |
| 5.        POST-ACCUM |
| 6.        [vertex operations];    // reduce phase for each selected node |
| 7. End; |

III. GRAPH BASED COMBINATION OF ITERATIVE METHODS AND VERTEX CONTRACTION FOR POWER FLOW ANALYSIS

A. *Bi-level PageRank Approach for Power Flow Analysis with Graph Computing*

*1) PageRank Using Graph Computing*: PageRank is an algorithm, firstly used by Google Search, to rank websites by calculating the web page importance. Its equation is as follows, indicating the PageRank of page $p_i$ at time point of *(t+1)*.

$$PR(p_i;t+1) = \frac{1-d}{N} + d \cdot (\sum_{p_j \in B_{p_i}} \frac{PR(p_j;t)}{L(p_j;t)}) \quad (1)$$

where *d* is the damping factor, generally around 0.85, $B_{P_i}$ is the set of pages linked to $p_i$, $L(p_j;t)$ is the number of out links on page $p_i$, and *N* is the number of pages.

In equation (1), two features of PageRank algorithm are presented: (a) the PageRank of each page is only determined by its neighboring PageRank and the number of its out links; (b) during each iteration, the PageRank computation for each page only uses the values obtained from the previous iteration. The former reveals that each page's PageRank can be locally calculated, and the second characteristic indicates that the PageRank algorithm can be implemented in parallel. Based on the description of graph computing, node-based parallel computing is applicable to PageRank.

*2) Jacobi Method for Power Flow Analysis Using Graph Computing*: The power flow equations are in (2). Using Jacobi method, the power flow iteration is shown in (3), which is like the PageRank algorithm. The voltage at each node is determined by its neighboring voltages, the connected line admittance, and its own information, like voltage magnitude, voltage angle, node admittance, power injection, etc. Besides, in each iteration, its calculation is determined by system states calculated from last iteration. Therefore, the power flow analysis is feasible in parallel computing. But the slow rate of convergence is an issue, even though it takes little memory and does not need to do matrix factorization and forward/backward substitution.

$$\dot{S}_i^* = P_i - jQ_i = \dot{V}_i^* \sum_{j=1}^{n} (Y_{ij} \cdot \dot{V}_j)$$

$$\rightarrow \dot{V}_i = \frac{1}{Y_{ii}} \left[ \frac{P_i - jQ_i}{\dot{V}_i^*} - \sum_{\substack{j=1 \\ j \neq i}}^{n} (Y_{ij} \cdot \dot{V}_j) \right] \quad (2)$$

$$\dot{V}_i^{(k+1)} = \frac{1}{Y_{ii}} \left[ \frac{P_i^{(k)} - jQ_i^{(k)}}{\dot{V}_i^{(k)*}} - \sum_{\substack{j=1 \\ j \neq i}}^{n} (Y_{ij} \cdot \dot{V}_j^{(k)}) \right] \quad (3)$$

*3) Bi-Level PageRank Method Using Graph Computing:* To improve its convergence-rate, this paper employs two strategies: (a) using damping factor, like PageRank algorithm; (b) separating nodes into two levels, borrowing the idea from Gauss-Seidel algorithm.

With the addition of the damping factor, equation (3) is developed into (4). Like the function in PageRank algorithm, damping factor is used to improve the convergence of power flow calculation. First, it could avoid a sink when zero-impedance branch exists. Look at (3), if there is a zero-impedance branch connected to node *i*, the values of $\sum_{\substack{j=1 \\ j \neq i}}^{n} Y_{ij} \dot{V}_j$ and $Y_{ii}$ are too large to reflect power changes in voltage update, leading to a voltage sink at node *i*, and a worse convergence. Also, if $Y_{ij}$ is too large, the condition number of the admittance matrix is too large to converge. In addition, using damping factor could help much reduce fluctuations around the real system state and improve the convergence-rate.

But, the convergence-rate is still slow, especially when a high precision is needed. This is determined by the algorithm itself. In each iteration, its computation only depends on results obtained from the previous iteration. Borrowing the idea of Gauss-Seidel to improve power flow convergence, this paper proposes a bi-level PageRank approach to improve the convergence and meanwhile maintain the capability of parallel computing in Jacobi method. The nodes are divided into two levels in the GDB, ensuring that most of nodes are not mutually connected within each level. So, based on equation (4), equation (5) is developed for the bi-level PageRank method based power flow analysis. The graph for the corresponding power system is also divided into two, graph *A* and graph *B*. Graph *A* is first updated using results from the previous iteration. Then nodes in graph *B* are updated using graph *A*'s results in current iteration and graph *B*'s information from last iteration [19].

$$\dot{V}_i^{(k+1)} = (1-d)\dot{V}_i^{(k)} + \frac{d}{Y_{ii}}\left[\frac{P_i^{(k)} - jQ_i^{(k)}}{\dot{V}_i^{(k)*}} - \sum_{\substack{j=1 \\ j \neq i}}^{n}(Y_{ij}\cdot\dot{V}_j^{(k)})\right] \quad (4)$$

$$\dot{V}_i^{(k+1)} = \quad (5)$$

$$\begin{cases} (1-d)\dot{V}_i^{(k)} + \frac{d}{Y_{ii}}\left[\frac{P_i^{(k)} - jQ_i^{(k)}}{\dot{V}_i^{(k)*}} - \sum_{\substack{j\in A \\ j\neq i}}(Y_{ij}\cdot\dot{V}_j^{(k)}) - \sum_{j\in B}(Y_{ij}\cdot\dot{V}_j^{(k)})\right] (i\in A) \\ (1-d)\dot{V}_i^{(k)} + \frac{d}{Y_{ii}}\left[\frac{P_i^{(k)} - jQ_i^{(k)}}{\dot{V}_i^{(k)*}} - \sum_{j\in A}(Y_{ij}\cdot\dot{V}_j^{(k+1)}) - \sum_{\substack{j\in B \\ j\neq i}}(Y_{ij}\cdot\dot{V}_j^{(k)})\right] (i\in B) \end{cases}$$

The computation procedure in graph computing is also displayed below.

**Graph Computing based Bi-Level PageRank Algorithm in Power Flow Analysis**

1. Initialize T0 = {all nodes}
2. T1 = SELECT s FROM T0:s-(edge:e)->t
3.    ACCUM
4.    [calculate off-diagonal elements in Ybus matrix],
5.    [sum up off-diagonal elements for each node in Ybus matrix]
6.    POST-ACCUM
7.    [complete diagonal elements calculation for Ybus matrix],
8.    [initialize system states];
9. while (Re{V} > threshold & Im{V}>threshold){
10. T2 = SELECT s FROM T1:s-(edge:e)->t
11.    ACCUM
12.    [calculate $Y_{ij}\dot{V}_j^{(k)}$ through edge operations]
13.    POST-ACCUM
14.    [update voltages via node operations],
15.    [update power mismatch and voltage changes from last iteration];}
16. End;

In Fig. 5, the convergence performance of the bi-level PageRank method for power flow analysis is presented, using a provincial FJ-1425 system. Beginning with flat-start, it converges very fast in the beginning when the voltage mismatch, respectively comparing magnitude and phase angle, is larger than 0.005. Then, the convergence curve becomes flat. Because of the existence of zero-impedance branch, the maximum bus power injection mismatch (MBPIM) is much higher than 0.05 per unit even when converged.

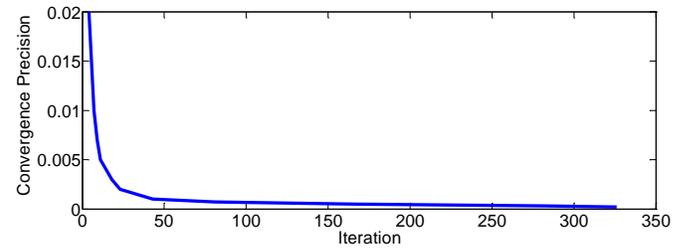

Fig. 5. Convergence of bi-level PageRank power flow analysis

### B. Diagonal Conjugate Gradient (DCG) Method for Power Flow Analysis Using Graph Computing

In mathematics, conjugate gradient algorithm is widely employed as an iterative approach to solve sparse and symmetric positive definite linear systems [11]. Besides, the convergence-rate of an iterative linear solver increases as the condition number of the coefficient matrix decreases [17]. Hence, a well-selected preconditioner *M* is beneficial for a linear system solving. If using the preconditioner *M* = *A*$^{-1}$, the solution can be obtained directly. However, there is an extra cost in the inverse operation of the coefficient matrix, *A*, and matrix multiplication, *A*$^{-1}$·*b*. So, the selection of the appropriate preconditioner is the trade-off between convergence improvement and matrix operation cost. In this paper, diagonal conjugate gradient (DCG) method is selected, because of its easy implementation and low time cost in graph database, and much better convergence-rate than conjugate gradient. It is widely employed to solve sparse symmetric positive definite linear systems [20]. The diagonal conjugate gradient algorithm is presented below.

**Diagonal Conjugate Gradient Algorithm**

1. Initialization: $x_0 = 0$, $r_0 = b - A\cdot x_0$, $z_0 = M^{-1}\cdot r_0$, $p_0 = z_0$
2. For *k = 1, 2, 3, ...*
3.    $\alpha_k = (r_{k-1}^T \cdot z_{k-1})/(p_{k-1}^T \cdot A \cdot p_{k-1})$;  Step Length
4.    $x_k = x_{k-1} + \alpha_k \cdot p_{k-1}$;  Approx. Solution
5.    $r_k = r_{k-1} - \alpha_k \cdot A \cdot p_{k-1}$;  Residual
6.    End if $r_k$ is sufficiently small, then exit loop.

7. $k++$;
8. $z_k := D^{-1} \cdot r_{k-1}$;
9. $\beta_k = (z_k^T \cdot r_k)/(z_{k-1}^T \cdot r_{k-1})$;    Improvement
10. $p_k = z_k + \beta_k \cdot p_{k-1}$;    Search Direction
11. End for-loop;

Based on the conjugate gradient algorithm and the feature of graph database, variables in DCG are categorized into local variables, neighboring variables, and global variables, as shown in Fig. 6. Local variables are available and can be directly updated using information from "local" vertices. The neighboring variable update needs information from vertices and edges, meaning it must do one-step graph traversal to retrieve information from connected edges and neighboring vertices. For the global variable update, it needs information from local and neighboring variables via a full graph traversal. After the decomposition and analysis, DCG could be implemented using node-based parallel graph computing.

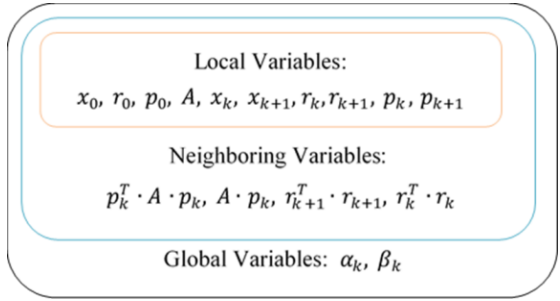

Fig. 6. Variables classification in the DCG approach

In this way, the implementation of DCG in graph computing is demonstrated below, where $.BIJ$ indicate the non-zero element in the matrix $A$, including the off-diagonal element for edge, $e.BIJ$, and the diagonal element for vertex, $s.BIJ$.

**Diagonal Conjugate Gradient Algorithm Implemented in Graph Computing**

1. Initialization: $x_0 = 0$, $r_0 = b - A \cdot x_0$, $z_0 = M^{-1} \cdot r_0$, $p_0 = z_0$
2. For $k = 1, 2, 3, ...$
3.    SELECT EDGE $e$ FROM VERTEX $s$ in T0 to VERTEX $t$
4.    WHERE statement // filter out disconnected edge
5.    ACCUM
6.       $s.TempAP += e.BIJ \cdot t.P_{k-1}$,
7.       $PAP += t.P_{k-1} \cdot e.BIJ \cdot t.P_{k-1}$
8.    POST-ACCUM
9.       $t.x_k = t.x_{k-1} + \alpha_k \cdot t.P_{k-1}$,
10.      $t.r_k = t.r_{k-1} - \alpha_k \cdot t.TempAP$,
11.      $t.z_k = t.r_k / t.BIJ$,
12.      $r_k^2 += t.r_k \cdot t.z_k$;
13. End the loop if $r_k$ is sufficiently small;
14. $\beta_k = r_k^2 / r_{k-1}^2$;
15. $k++$;
16. SELECT VERTEX s in T0
17. POST-ACCUM
18.    $s.TempAP = 0$,
19.    $s.P_k = s.z_k + \beta_k \cdot s.P_{k-1}$,
20.    $s.TempAP += s.BIJ \cdot t.P_k$,
21.    $PAP += s.BIJ \cdot s.P_k$;
22. End for-loop;

In this paper, DCG method is applied to power system analysis by making use of its high parallelism. The linearized power flow equations are in equations (6) and (7).

$$\Delta P/V = B' \cdot \Delta \theta \quad (6)$$
$$\Delta Q/V = B'' \cdot \Delta V \quad (7)$$

In addition, the convergence of DCG highly depends on the initial input $x_0$ of the linear system $A \cdot x = b$. Fig.7 shows the convergence of DCG method. Compared with Fig. 5, it took ~1000 iterations to reach the precision of 0.005 for voltage mismatch, then it only costs approximately 200 iterations to converge and the maximum bus power injection MBPIM is less than 0.05 per unit. It clearly shows that if $x_0$ is close to the final solution, it will converge quickly. In the Section III.D, DCG will be combined with bi-level PageRank approach in the power flow analysis. Then, DCG could employ the output of bi-level PageRank approach as the input of DCG power flow analysis, and the convergence performance is largely improved.

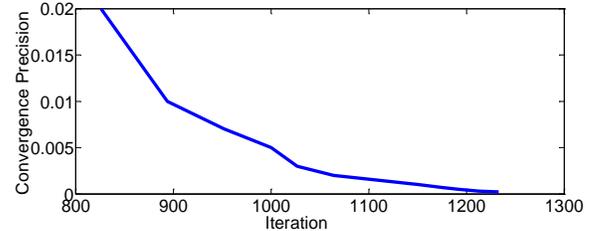

Fig. 7. Convergence of DCG power flow analysis

*C. Vertex Contraction for Power Flow Analysis Convergence Improvement*

Before combining bi-level PageRank approach and DCG algorithm to improve the power flow analysis performance, one more step is needed to reduce, even avoid, zero-impedance branch impact on the convergence-rate. Because of the existence of zero-impedance branches, diagonal elements in the coefficient matrix are not dominant, worsening the convergence. In addition, as seen in (1), if zero-impedance branch exists, the branch admittance, $Y_{ii}$, is very large, attenuating the influence of bus power injection. So, it is difficult to guarantee that the value of maximum bus power injection mismatch is within an acceptable range in (1). In this section, the approach of vertex contraction (VC) is employed to eliminate zero-impedance branch for power flow analysis, as shown in Fig. 8. In this way, the condition number of the coefficient matrix in the power flow equation is much reduced to largely improve the convergence-rate for iterative methods. This will not affect the final solution, since the bus states on both sides of each zero-impedance branch have few deviations unless nonnegligible branch is neglected.

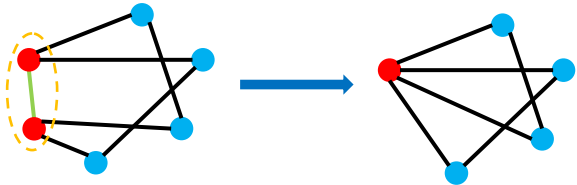

Fig. 8. Vertex contraction in graph database

### D. Graph based Combination of Iterative Methods and Vertex Contraction Approach for Power Flow Analysis

As shown in Fig. 5 and Fig. 7, bi-level PageRank method converges fast in the beginning and DCG can converge quickly with a good initial start. Besides, with the use of VC, the impact of zero-impedance branch is minimized. So, this paper proposes to first employ VC for network preprocessing, use bi-level PageRank to achieve a good initial start for DCG and, at last, reach the final solution using DCG algorithm. It is able to improve the convergence-rate without sacrificing the computation accuracy. Different methods are tested and compared in Section IV.

## IV. CASE STUDY

### A. Performance Testing

In this section, a practical power system in Fujian province with 1425 buses are used as the testing case. As shown in Table I, five different approaches for power flow analysis are implemented in FJ-1425 system. The comparison of computation performance, including both computation speed and calculation accuracy, is presented. It can be clearly seen that, with the help of VC, the impact of zero-impedance branches is minimized and the convergence-rate is largely improved. This is because the condition number is much reduced after the elimination of the zero-impedance branch. In addition, it displays that only Bi-level PageRank method is not effective for a large system, even though its convergence-rate is much better than DCG, which needs a good initial start. Regarding the fifth approach, which is a combination of iterative methods and VC, its computation time is comparable with Bi-level PageRank and much less than DCG and the method of VC+DCG. Furthermore, the calculation accuracy is much better than Bi-level PageRank, and the same as DCG and the approach of VC+DCG.

TABLE I
PERFORMANCE OF DIFFERENT GRAPH COMPUTING BASED METHODS FOR POWER FLOW ANALYSIS IN FJ-1425 SYSTEM

| Approach | Convergence Precision | Number of Iteration | Time (ms) | MBPIM < 5E-2 p.u. |
|---|---|---|---|---|
| Bi-Level PageRank | $\Delta\|V_r\| < $ 2E-4<br>$\Delta\|V_a\| < $ 2E-4 | 365 | 364.58 | No |
| VC based Bi-Level PageRank | $\Delta\|V_r\| < $ 2E-4<br>$\Delta\|V_a\| < $ 2E-4 | 215 | 272.98 | No |
| DCG | $\Delta P < $ 5E-2<br>$\Delta Q < $ 5E-2 | 1123 | 1245.10 | Yes |
| VC + DCG | $\Delta P < $ 5E-2<br>$\Delta Q < $ 5E-2 | 625 | 774.51 | Yes |
| VC + Bi-Level PageRank + DCG | ① Bi-Level PageRank: $\Delta\|V_r\| < $ 1E-2 $\Delta\|V_a\| < $ 1E-2;<br>② DCG: $\Delta P < $ 5E-2 $\Delta Q < $ 5E-2 | ① Bi-Level PageRank: 8<br>② DCG: 318<br>Total: 326 | 370.77 | Yes |

### B. Discussion

Assuming that system state changes gradually in continuous time series, then the problem solution could be converged very fast if using the system state at the last time point as the start point. Besides, using the method of VC + Bi-level PageRank + DCG has no LU factorization and complicated matrix manipulation, saving large amount of time.

## V. CONCLUSION

In this paper, a combinative approach is implemented for fast power flow analysis. It consists of vertex contraction, Bi-level PageRank and DCG. It is verified that using vertex contraction can help convergence improvement by avoiding zero-impedance branch, Bi-level PageRank is able to converge fast in the beginning to provide a good initial start for DCG, and then DCG can converge quickly without sacrificing the computation accuracy.

### REFERENCES

[1] U.S. Department of Energy, "2015 Quadrennial Technology Review," 2015. [Online]. Available: https://www.energy.gov/sites/prod/files/2017/03/f34/quadrennial-technology-review-2015_1.pdf.
[2] S. Zhao and C. Singh, "Studying the Reliability Implications of Line Switching Operations," *IEEE Trans. Power Syst.*, vol. 32, no. 6, pp. 4614–4625, 2017.

2018 International Conference on Power System Technology (POWERCON 2018)     Guangzhou, 6-8 Nov. 2018